\documentclass[a4paper,reqno]{amsart}

\usepackage[all]{xy}           
\usepackage{amssymb}           
\usepackage{hyperref}
\usepackage{eucal}
\usepackage{nicefrac}
\numberwithin{equation}{section}

\newtheorem{definition}{Definition}[section]

\newtheorem{theorem}[definition]{Theorem}

\newtheorem{corollary}[definition]{Corollary}
\newtheorem{remarkth}[definition]{Remark}

\newenvironment{remark}{\begin{remarkth}\upshape}{\hfill$\diamond$\end{remarkth}}

\renewcommand{\emph}[1]{{\bfseries\itshape{#1}}}

\newcommand{\R}{\mathbb{R}}      

\newcommand{\ak}{1\leq \alpha \leq k}
\newcommand{\n}{1\leq i \leq n}

\newcommand{\ds}{\displaystyle}

\def\derpar#1#2{\ds\frac{\partial{#1}}{\partial{#2}}}
\def\vf{\mathfrak{X}}
\makeatletter
\newcommand\prol{\@ifstar{\@proldf}{\@prolpf}}  
\def\@prolpf{\@ifnextchar[{\@prolpf@wrt}{\@prolpf@}}
\def\@prolpf@wrt[#1]#2{\@ifnextchar[{\@prolpf@wrt@at{#1}{#2}}{\@prolpf@wrt@{#1}{#2}}}
\def\@prolpf@wrt@at#1#2[#3]{\prolsymbol^{#1}_{#3}#2}
\def\@prolpf@wrt@#1#2{\prolsymbol^{#1}#2}
\def\@prolpf@#1{\@ifnextchar[{\@prolpf@at{#1}}{\@prolpf@@{#1}}}
\def\@prolpf@at#1[#2]{\prolsymbol_{#2}#1}
\def\@prolpf@@#1{\prolsymbol#1}
\def\@proldf{\@ifnextchar[{\@proldf@wrt}{\@proldf@}}
\def\@proldf@wrt[#1]#2{\@ifnextchar[{\@proldf@wrt@at{#1}{#2}}{\@proldf@wrt@{#1}{#2}}}
\def\@proldf@wrt@at#1#2[#3]{\prolsymbol^{*#1}_{#3}#2}
\def\@proldf@wrt@#1#2{\prolsymbol^{*#1}#2}
\def\@proldf@#1{\@ifnextchar[{\@proldf@at{#1}}{\@proldf@@{#1}}}
\def\@proldf@at#1[#2]{\prolsymbol^*_{#2}#1}
\def\@proldf@@#1{\prolsymbol^*#1}
\def\prolsymbol{\mathcal{T}}
\makeatother

\def\r{\ensuremath{\mathbb{R}}}
\def\rk{{\mathbb R}^{k}}

\def\tkqh{(T^1_k)^*Q}

\def\rktkqh{\rk\times (T^1_k)^*Q}
\def\rkq{\rk \times Q}

\begin{document}

\title[Hamilton-Jacobi theory in $k$-cosymplectic field theories]{Hamilton-Jacobi theory in $k$-cosymplectic field theories}

\author[M. de Le\'on]{M. de Le\'on}
\address{M. de Le\'on:
Instituto de Ciencias Matem\'aticas (CSIC-UAM-UC3M-UCM),
Consejo Superior de Investigaciones Cient\'{\i}ficas, Serrano 123, 28006
Madrid, Spain} \email{mdeleon@icmat.es}


\author[S. Vilari\~no]{S. Vilari\~no}
\address{S. Vilari\~no:
Centro Universitario de la Defensa $\&$ I.U.M.A.,
    Carretera de Huesca s/n,
    50090 Zaragoza, Spain}
    \email{silviavf@unizar.es}

\today{ }


 \subjclass[2000]{}

\begin{abstract}
 In this paper we extend the geometric formalism of the Hamilton-Jacobi
 theory for time dependent Mechanics to the case of classical field theories in the $k$-cosymplectic framework.

\end{abstract}


 \maketitle

\tableofcontents

\section{Introduction}

The usefulness of Hamilton-Jacobi theory in Classical Mechanics
is well-known, giving an alternative  procedure to study and, in some  cases, to solve the evolution equations \cite{am}.
The use of symplectic geometry in the study of Classical Mechanics
has permitted to connect the Hamilton-Jacobi theory with the theory
of Lagrangian submanifolds and generating functions \cite{BLM-2012}.

At the beginning of the 1900s an analog of Hamilton-Jacobi equation for field theory has been developed
\cite{rund}, but it has not been proved to be as powerful as the theory which is available for mechanics \cite{bertin,bruno,pau1,pau2,rosen,vita}.

There are several recent  attempts to extend the Hamilton-Jacobi theory  for classical field theories in a geometrical setting. For instance in the framework of the so-called multisymplectic formalism \cite{lmm2,{pau1},{pau2}} (see also \cite{CIL1,kt} for a general description of the multisymplectic setting) or in the $k$-symplectic formalism in \cite{LMMSV-2010} (see  \cite{MRS-2004,RRSV-2011} for a discussion of the relationship between both formulations, see also \cite{ultimo}). Our method is based in that developed by J.F. Cari\~{n}ena {\it et al.} for Classical Mechanics \cite{CGMMR,{pepin2}} (see also \cite{lmm1,lmm3}).

In the context of Classical Field Theories, the Hamiltonian is a function $H=H(x^\alpha, q^i, p^\alpha_i)$, where $(x^\alpha)$ are coordinates in the space-time, $(q^i)$ represent the field coordinates and $(p^\alpha_i)$ are the conjugate momenta.  In this context, the Hamilton-Jacobi equation is \cite{rund}
\begin{equation}\label{HJCFT}
    \derpar{W^\alpha}{x^\alpha} + H\Big(x^\beta, q^i, \derpar{W^\alpha}{q^i}\Big)=0
\end{equation}
where $W^1,\ldots, W^k\colon \rkq\to \r$.

The aim of this paper is to extend the Hamilton-Jacobi theory to field theories just in
the context of $k$-cosymplectic manifolds \cite{merino1,merino2,merino3}. The ``dynamics'' for a given Hamiltonian function $H$ is interpreted
as a family of vector fields (a $k$-vector field) on the phase space $\rk\times(T^1_k)^*Q$ (or in general on a $k$-cosymplectic manifold $(M,\eta^\alpha, \omega^\alpha, V)$).

The paper is structured as follows. In Sec. \ref{k-cosmp}, we recall the notion of $k$-vector
field and their integral sections and give a briefly description of the $k$-cosymplectic formalism. In Sec. \ref{HJsection} we discuss  the Hamilton-Jacobi equation in the $k$-cosymplectic context.  Finally, an
example is discussed in Sec. \ref{example}, with the aim to show how the method works.

We shall also adopt the convention that a repeated index implies summation over the range of the index, but, in some cases, to avoid confusions we will explicitly include the summation symbol.

\section{The $k$-cosymplectic formalism}\label{k-cosmp}
The $k$-cosymplectic formalisms \cite{merino1,merino3} is one of the simplest geometric frameworks for describing first order
classical field theories (see \cite{aw1} for the $k$-symplectic case). It is the generalization to field theories of the standard cosymplectic
formalism for nonautonomous mechanics  and it describes field theories involving the space-time coordinates
on the Lagrangian and on the Hamiltonian cases. The foundations of the $k$-cosymplectic
formalism are the k-cosymplectic manifolds.
In this section we briefly recall this formalism.

\subsection{Geometric preliminaries: k-vector fields and integral sections.}

In this section we briefly recall some well-known facts about
tangent bundles of $k^1$-velocities (we refer the reader to
\cite{mt1,mt2,arxive,mor,MRS-2004, RRSV-2011} for more details).

Let $\tau_M : TM \longrightarrow M$ be the tangent bundle of $M$.
Let us denote by $T^1_kM$ the Whitney sum $TM \oplus
\stackrel{k}{\dots} \oplus TM$ of $k$ copies of $TM$, with
projection $\tau: T^1_kM \longrightarrow M$, $\tau
({v_1}_{p},\dots , {v_k}_{p})=p$, where ${v_\alpha}_{p}\in T_{p}M$,
$1\leq \alpha \leq k$. $T^1_kM$ can be identified with the manifold
$J^1_0(\R^k,M)$ of the $k^1$-velocities of $M$, that is, $1$-jets of
maps $\eta : \R^k \longrightarrow M$ with source at $0 \in \R^k$,
say
\[
\begin{array}{ccc}
J^1_0(\R^k,M) & \equiv & TM \oplus \stackrel{k}{\dots} \oplus TM \\
j^1_{0, p} \eta & \equiv & ({v_1}_p, \dots , {v_k}_p)
\end{array}
\]
where $x=\eta (0)$, and ${v_\alpha}_p = T\eta
(0)(\ds\frac{\partial}{\partial x^\alpha} \Big\vert_{0})$. Here
$(x^1,\ldots, x^k)$ denote the standard coordinates on $\R^k$.
$T^1_kM$ is called the {\it tangent bundle of $k^1$-velocities} of
$M$ or simply $k$-tangent bundle for short, see \cite{mor}.

Denote by $(x^i , v^i)$ the fibred coordinates in $TM$ from local
coordinates $(x^i)$ on $M$. Then we have fibred coordinates $(x^i ,
v_\alpha^i)$, $1\leq i \leq m,\, 1 \leq \alpha \leq k$, on $T^1_kM$, where
$m=\dim M$.

\begin{definition}\label{kvector}
A section $ X  : M \longrightarrow T^1_kM$ of the projection
$\tau$ will be called a {\rm $k$-vector field on $M$}.
\end{definition}

Since $T^{1}_{k}M$ is  the Whitney sum $TM\oplus \stackrel{k}{\dots}
\oplus TM$ of $k$ copies of $TM$, we deduce given a
$k$-vector field $ X $ is equivalent to giving a family of $k$
vector fields $X_{1}, \dots, X_{k}$ on $M$ by projecting $X$
onto each factor. For this reason we will denote a $k$-vector field
by $(X_1, \dots, X_k)$.

\begin{definition}
\label{integsect} {\rm An integral section} of the $k$-vector field $
X=(X_{1}, \dots,X_{k})$, passing through a point $p \in M$, is a
map $\psi \colon U_0 \subset \R^k \longrightarrow M$, defined on
some neighborhood $U_0$ of $0 \in \R^k$, such that
$$
\psi(0) = p, \, \, T\psi\left(\ds\frac{\partial}{\partial
x^\alpha}\Big\vert_{x}\right)=X_\alpha(\psi (x))
 , \quad \mbox{\rm for every $x \in U_0$, $1\leq \alpha \leq k$}
$$
or, equivalently,  $\psi$ satisfies that
$X\circ\psi=\psi^{(1)}$, being  $\psi^{(1)}$ is the first
prolongation of $\psi$  to $T^1_kM$ defined by
$$
\begin{array}{rccl}\label{1prolong}
\psi^{(1)} : & U_0\subset \R^k & \longrightarrow & T^1_kM
\\\noalign{\medskip}
 & x & \longrightarrow & \psi^{(1)} (x)= j^1_0 \psi_x \;,
 \end{array}
$$
where $\psi_x(s) = \psi(x+s)$.

A $k$-vector field $X=(X_1,\ldots , X_k)$ on $M$ is said to be
{\rm integrable} if there is an integral section passing through every
point of $M$.
\end{definition}

In local coordinates, we have
\begin{equation} \label{localfi11}
\psi^{(1)}(x^1, \dots, x^k)=\left( \psi^i (x^1, \dots, x^k),
\ds\frac{\partial\psi^i}{\partial x^\alpha} (x^1, \dots, x^k)\right),
\end{equation}
and then $\psi$ is an integral section of $(X_1, \dots, X_k)$ if and
only if the following equations hold:
\begin{equation}\label{condintsect}
\ds\frac{\partial \psi^i}{\partial x^\alpha}= X_\alpha^i \circ \psi\, \quad
1\leq \alpha \leq k,\; 1\leq i\leq m\;,
\end{equation}
being $X_\alpha=X_\alpha^i\ds\frac{\partial}{\partial q^i}$.

Notice that, in case $k=1$, Definition \ref{1prolong} coincides with
the definition of integral curve of a vector field.

\subsection{$k$-cosymplectic manifolds}

Let $Q$ be a differentiable manifold, $\dim Q = n$, and
$\pi: T^*Q \to Q$ its cotangent bundle.
Denote by $(T^1_k)^*Q= T^*Q \oplus \stackrel{k}{\dots}
\oplus T^*Q$, the Whitney sum of $k$ copies of $T^*Q$. The
manifold $(T^1_k)^*Q$ can be identified with the manifold
$J^1(Q,\rk)_0$ of 1-jets of mappings from $Q$ to $\rk$ with target
at $0\in \rk$, the diffeomorphism is given by
\[
\begin{array}{ccc}
J^1(Q,\r^k)_0 & \equiv & T^*Q \oplus \stackrel{k}{\dots} \oplus T^*Q \\
j^1_{q,0}\sigma  & \equiv & (d\sigma^1(q), \dots ,d\sigma^k(q))\ ,
\end{array}
\]
where $\sigma^\alpha= \pi^\alpha \circ \sigma:Q \longrightarrow \r$ is the
$\alpha^{th}$ component of $\sigma$, and  $\pi^\alpha:\r^k \to \r$ is the
canonical projection onto the $\alpha^{th}$ component, for $ \alpha = 1, \ldots ,
k$. $(T^1_k)^*Q$ is called {\sl the cotangent bundle of
$k^1$-covelocities of the manifold $Q$}.

The manifold $J^1\pi_{Q}$ of  1-jets of sections of the trivial
bundle $\pi_{Q}:\rk \times Q \to Q$ is diffeomorphic to $\rk \times
(T^1_k)^*Q$, via the diffeomorphism given by
\[
\begin{array}{rcl}
J^1\pi_{Q} & \to & \rk   \times (T^1_k)^*Q \\
\noalign{\medskip} j^1_q\phi= j^1_q(\phi_{\rk},Id_{Q})  & \mapsto &
(\phi_{\rk}(q), \nu^1_q, \ldots ,\nu^k_q) \ ,
\end{array}
\]
where $\phi_{\rk}: Q \stackrel{\phi}{\to}  \rkq
\stackrel{\pi_{\rk}}{\to}\rk $, and $\nu_q^\alpha = d \phi_{\rk}^\alpha(q)$,
$1\leq \alpha \leq k.$

Throughout the paper, we use the following notation for the canonical projections
\[\xymatrix@C=13mm{\rk\times (T^1_k)^*Q\ar[r]^-{({\pi}_Q)_{1,\,0}}\ar[dr]_-{({\pi}_Q)_1}
& \rkq\ar[d]^-{{\pi}_Q}\\
 & Q
}\]
  where
$$\pi_Q(x,q)=q, \quad (\pi_Q)_{1,0}(x,\nu^1_q, \ldots
,\nu^k_q)=(x,q), \quad (\pi_Q)_1(x,\nu^1_q, \ldots
,\nu^k_q)=q \, ,$$ with $x\in \rk $, $q\in Q$ and $(\nu^1_q,
\ldots ,\nu^k_q)\in (T^1_k)^*Q$.

If $(q^i)$ are local coordinates on $U \subseteq Q$, then the
induced local coordinates $(q^i , p_i)$, $1\leq i \leq n$, on
$(\tau^*_Q)^{-1}(U)=T^*U\subset T^*Q$, are given by
$$
q^i(\nu_q)=q^i(q), \quad p_i(\nu_q)= \nu_q \left(
\frac{\partial}{\partial q^i}\Big\vert_q \right)\, ,
$$
with $\nu_q\in T^*Q$,
and the induced local coordinates  $(x^\alpha,q^i ,
p^\alpha_i)$ on $[(\pi_Q)_1]^{-1}(U)=\rk \times (T^1_k)^*U$ are given
by
$$
x^\alpha(x,\nu^1_q, \ldots ,\nu^k_q) = x^\alpha, \;
q^i(x,\nu^1_q, \ldots ,\nu^k_q) = q^i(q), \;
p_\alpha^i(x,\nu^1_q, \ldots ,\nu^k_q) =
\nu^\alpha_q\left(\ds\frac{\partial}{\partial q^i}\Big\vert_q
\right),
$$
for $1\leq i\leq n$ and $1\leq \alpha\leq k$.

On $\rk\times (T^1_k)^*Q$, we consider the differential forms
$$
 \eta^\alpha=dx^\alpha=(\pi^\alpha_1)^*dx\, , \quad \theta^\alpha=
(\pi^\alpha_2)^*\theta\, , \quad \omega^\alpha=
(\pi^\alpha_2)^*\omega\, ,
$$
where $\pi^\alpha_1:\rk \times (T^1_k)^*Q \rightarrow \r$ and
$\pi^\alpha_2:\rk \times (T^1_k)^*Q \rightarrow T^*Q$ are the
projections defined by
$$
\pi^\alpha_1(x,\nu^1_q, \ldots ,\nu^k_q)=x^\alpha ,\quad
\pi^\alpha_2(x,\nu^1_q, \ldots ,\nu^k_q)=\nu^\alpha_q\, ,
$$
$\omega=-d\theta=dq^i \wedge dp_i$ is the canonical symplectic form
on $T^*Q$ and $\theta=p_i \, dq^i$ is the Liouville $1$-form on
$T^*Q$. Obviously $\omega^\alpha = -d\theta^\alpha$.

In local coordinates we have
\begin{equation}\label{locexp}
    \eta^\alpha=dx^\alpha   ,  \quad  \theta^\alpha = \displaystyle \, p^\alpha_i
dq^i   ,  \quad \omega^\alpha = \displaystyle  dq^i \wedge dp^\alpha_i\, .
\end{equation}

Moreover, let
\begin{equation}\label{distr} V=ker\, \left( \, (\pi_Q)_{1,0}
\right)_*=\left\langle\frac{\displaystyle\partial}
{\displaystyle\partial p^1_i}, \dots,
\frac{\displaystyle\partial}{\displaystyle\partial
p^k_i}\right\rangle_{i=1,\ldots , n}
\end{equation}
be the vertical distribution of the bundle
$(\pi_{Q})_{1,0}:\rk\times (T^1_k)^*Q \to \rkq$.

A simple inspection of the expressions in local coordinates,
(\ref{locexp}) and (\ref{distr}) show that the forms $\eta^\alpha$ and $\omega^\alpha$ are
closed, and the following relations hold
\begin{enumerate}
\item
 $\eta^1\wedge\dots\wedge \eta^k\neq 0$,\quad
$(\eta^\alpha)\vert_{V}=0,\quad (\omega^\alpha)\vert_{V\times
V}=0,$
\item
 $(\ds {\cap_{\alpha=1}^{k}} \ker \eta^\alpha) \cap (\ds
{\cap_{\alpha=1}^{k}} \ker \omega^\alpha)=\{0\}$, \quad $dim(\ds
{\cap_{\alpha=1}^{k}} \ker \omega^\alpha)=k,$
\end{enumerate}

Inspired by the above  geometrical model we introduce the following  (see \cite{merino3}),
\begin{definition}\label{deest}
Let $M$ be a differentiable manifold  of dimension $k(n+1)+n$. A
{\rm $k$--cosymplectic structure} on $M$ is a family
$(\eta^\alpha,\omega^\alpha,V;1\leq \alpha\leq k)$, where each $\eta^\alpha$ is a closed
$1$-form, each $\omega^\alpha$ is a closed $2$-form and $V$ is an
integrable  $nk$-dimensional distribution on $M$ satisfying $(i)$ and
$(ii)$. $M$ is said to be a {\rm $k$--cosymplectic manifold}.
\end{definition}

The following theorem has been proved in \cite{merino3}:

\begin{theorem}[{\bf Darboux theorem}]\label{Darboux}
If $M$ is a $k$--cosymplectic manifold, then around each
  point of $M$ there exist local coordinates
$(x^\alpha,q^i,p^\alpha_i;1\leq \alpha\leq k, 1\leq i \leq n)$ such that
$$
\eta^\alpha=dx^\alpha,\quad \omega^\alpha=dq^i\wedge dp^\alpha_i, \quad
V=\left\langle\frac{\displaystyle\partial} {\displaystyle\partial
p^1_i}, \dots, \frac{\displaystyle\partial}{\displaystyle\partial
p^k_i}\right\rangle_{i=1,\ldots , n}.
$$
\end{theorem}

These coordinates will be called Darboux coordinates.
The canonical model for these geometrical structures is   $(\rk
\times (T^1_k)^*Q,\eta^\alpha,\omega^\alpha,V)$.

\subsection{$k$-cosymplectic Hamiltonian field theory.}

In this section we introduce the $k$-cosymplectic description of the Hamilton-De Donder-Weyl equations
\begin{equation}\label{HEQ} \frac{\ds \partial \psi^i}{\ds \partial x^\alpha}\Big\vert_{x}\,=
    \, \frac{\ds \partial H}{\ds \partial p^\alpha_i}\Big\vert_{ \psi(x)}\,
   ,\quad \ds\sum_{\alpha=1}^k\frac{\ds \partial \psi^\alpha_i}{\ds \partial x^\alpha}\Big\vert_{x}\,=
    \,- \frac{\ds \partial H}{\ds \partial q^i}\Big\vert_{\psi(x)}\,,\end{equation} where locally $\psi(x)=(x,\psi^i(x), \psi^\alpha_i(x))$. This approach was firstly introduced by  M.
de Le\'{o}n {\it et al.} \cite{merino3}. We will consider the general case on an arbitrary $k$-cosymplectic manifold $M$ but
everything can be particularize for the case of $M = \R^k \times (T_k^1)^*Q$.

\begin{definition}
    Let $(M,\eta^\alpha,\omega^\alpha,V)$ be a $k$-cosymplectic manifold and $H\colon M\to \r$ be a Hamiltonian function. The family $(M,\eta^\alpha, \omega^\alpha, H)$ is called {\rm $k$-cosymplectic Hamiltonian system}.
\end{definition}
\begin{theorem}\label{fhkc}Let $(M,\eta^\alpha, \omega^\alpha, H)$ a $k$-cosymplectic Hamiltonian system and $X=(X_1,\ldots, X_k)$ a $k$-vector field on  $M$ solution to the system of equations
\begin{equation}\label{geonah}
\begin{array}{l}
\eta^\alpha(X_\beta)=\delta^\alpha_\beta, \quad 1\leq \alpha,\beta\leq k\\
\noalign{\medskip}\displaystyle \sum_{\alpha=1}^k \,
 \imath_{X_\alpha}\omega^\alpha =
dH-\displaystyle\sum_{\alpha=1}^k R_\alpha(H)\eta^\alpha \, \, ,
\end{array}
\end{equation}
where $R_1,\ldots, R_k$ are the Reeb vector fields associated with the $k$-cosymplectic structure on $M$ which are characterized by the conditions
$$
\imath_{R_\alpha}\eta^\beta=\delta^\beta_\alpha  ,\quad \imath_{R_\alpha}\omega^\beta=0 \ .
$$

 If $\psi:\rk\to M,\,\psi(x)=(x^\alpha,\psi^i(x),\psi^\alpha_i(x))$ is an integral section of the $k$-vector field $X$, then $\psi$ is a solution of the Hamilton-De Donder-Weyl equations (\ref{HEQ}).
\end{theorem}

\proof   Let $X=(X_1,\dots,X_k)$ be a $k$-vector on   $M$ solution to  (\ref{geonah}). In Darboux coordinates each component $X_\alpha$ of the $k$-vector field    $X=(X_1,\dots,X_k)$   has the following local expression
  $$X_\alpha=(X_\alpha)_\beta\frac{\partial}{\ds\partial x^\beta} + (X_\alpha)^i\frac{\partial}{\ds\partial q^i}
  +(X_\alpha)^\beta_i\frac{\partial}{\ds\partial p^\beta_i}\,.$$

  Now, since
  \[
    dH=\derpar{H}{x^\alpha}dx^\alpha + \derpar{H}{q^i}dq^i + \derpar{H}{p^\alpha_i}dp^\alpha_i\,,
  \]
  and
  \[
    \eta^\alpha=dx^\alpha,\quad \omega^\alpha=dq^i\wedge dp^\alpha_i,\quad R_\alpha=\derpar{}{x^\alpha}\,,
  \]
  with $\ak$, we deduce that  equation (\ref{geonah}) is locally equivalent to the following local equations
  \begin{equation}\label{partials h X}
(X_\alpha)_\beta=\delta_\alpha^\beta, \quad \quad \ds\frac{\partial H}{\partial
p^\alpha_i}=(X_\alpha)^i, \quad \ds\frac{\partial H}{\partial q^i}=
-\ds\sum_{\alpha=1}^k(X_\alpha)^\alpha_i\,.
\end{equation}

  Let us suppose now that  $X=(X_1,\dots,X_k)$ is integrable and
  \[
    \begin{array}{rccl}
        \psi\colon &\rk& \to & M\\\noalign{\medskip}
         & x&\mapsto & \psi(x)=(x^\alpha,\psi^i(x),\psi^\alpha_i(x))
    \end{array}
  \]
  is an integral section of $ X $ then (see (\ref{condintsect}))
\begin{equation}\label{112}
(X_\alpha)_\beta\circ \psi=\delta_\alpha^\beta, \quad
\ds\frac{\partial \psi^i}{\partial x^\alpha}=(X_\alpha)^i\circ\psi, \quad
\ds\frac{\partial \psi^\beta_i}{\partial x^\alpha}=(X_\alpha)^\beta_i\circ\psi \,
.\end{equation}

 Therefore, from  (\ref{partials h X}) and (\ref{112}) we obtain that
$\psi(x)=(x,\psi^i(x),\psi^\alpha_i(x))$ is a solution of the following equations
$$
\frac{\ds\partial H}{\ds\partial q^i}\Big\vert_{\psi(x)}=
-\ds\sum_{\alpha=1}^k\frac{\ds\partial\psi^\alpha_i} {\ds\partial
x^\alpha}\Big\vert_{x} \;, \quad \frac{\ds\partial H} {\ds\partial
p^\alpha_i}\Big\vert_{\psi(x)}= \frac{\ds\partial\psi^i}{\ds\partial
x^\alpha}\Big\vert_{x},
$$
where $\n$ and $\ak$,
that is, $\psi$ is a solution to the Hamilton-De Donder-Weyl equations (\ref{HEQ}).
\qed

Consequently, the equations (\ref{geonah}) can be considered as a geometric version of the Hamilton-De Donder-Weyl field equations. From now, we will call these equations (\ref{geonah}) as {\rm $k$-cosymplectic Hamiltonian equations}.

\begin{definition}
    A $k$-vector field $X=(X_1,\ldots, X_k)$ is called a {\rm $k$-cosymplectic Hamiltonian $k$-vector field} for a $k$-cosymplectic Hamiltonian system $(M,\eta^\alpha,\omega^\alpha,$ $ H)$ if $X$ is a solution of (\ref{geonah}).
    We denote by $\vf^k_H(M)$ the set of $k$-vector fields which are solution of (\ref{geonah}).
\end{definition}

\begin{remark} \label{existence kcos sol}
We will discuss here about the existence and uniqueness of solutions
of equations (\ref{geonah}).

First of all, we shall prove the existence of geometric solutions.

Let $(M,\eta^\alpha,\omega^\alpha,V)$ be a
$k$-cosymplectic manifold; then, we can define the vector bundle morphism
\begin{equation}\label{sharp}
\begin{array}{rccl}
\omega^\sharp: & T^1_kM & \longrightarrow & T^*M  \\
\noalign{\medskip}
 & (X_1,\dots,X_k) & \mapsto &
  \displaystyle \sum_{\alpha=1}^k \,
\imath_{X_\alpha}\omega^\alpha+ \eta^\alpha(X_\alpha)\eta^\alpha \, .
\end{array}
\end{equation}
and, denoting by $ \mathcal{ M}_k(C^\infty(M))$ the space of matrices of
order $k$ whose entries are functions on $M$, we also have the vector bundle
morphism
\begin{equation}\label{etasharp}
\begin{array}{rccl}
\eta^{\sharp}: & T^1_kM & \longrightarrow & \mathcal{ M}_k(C^\infty(M))  \\
\noalign{\medskip}
 & (X_1,\dots,X_k) & \mapsto &
\eta^{\sharp}(X_1,\dots,X_k) =  (\eta^\alpha(X_\beta))\, .
\end{array}
\end{equation}

From the local conditions (\ref{partials h X}) we can define in a
neighborhood of each point $x\in M$ a $k$-vector field that
satisfies (\ref{geonah}). For example we can put
$$(X_\alpha)_\beta= \delta^\beta_\alpha \;, \quad (X_1)^1_i=\ds\frac{\partial H}{\partial q^i}\;, \quad
(X_\alpha)^\beta_i=0 \,\,\,\mbox{for $\alpha\neq 1\neq \beta$} \;, \quad (X_\alpha)^i=
\ds\frac{\partial H}{\partial p^\alpha_i}\, .$$
Now one can construct a
global $k$-vector field, which is a solution of (\ref{geonah}), by
using a partition of unity in the manifold $M$ (see \cite{merino1,merino3} for more details).

It should be noticed that, in general, equations (\ref{geonah}) do
not have a unique solution. In fact, the solutions of (\ref{geonah}) are given by
$(X_1,\dots,X_k)+(\ker\omega^{\sharp}\cap\ker\eta^{\sharp})$ for a particular solution $(X_1,\dots,X_k)$.

Let us observe that given a $k$-vector field $Y=(Y_1,\ldots, Y_k)$ the condition $Y\in \ker\omega^{\sharp}\cap\ker\eta^{\sharp}$ is locally equivalent to
\begin{equation}\label{kerkco}
(Y_\beta)_\alpha=0,\quad Y^i_\beta= 0, \quad \ds\sum_{\alpha=1}^k(Y_\alpha)^\alpha_i=0\,.
\end{equation}

\end{remark}

\begin{remark}
 In the case $k=1$ with $M=\r\times T^*Q$ the equations
(\ref{geonah}) reduces to the equations of the non-autonomous Hamiltonian Mechanics.
\end{remark}

\section{The Hamilton-Jacobi equation}\label{HJsection}

There are several attempts to extend the Hamilton-Jacobi theory  for classical field theories. In \cite{LMMSV-2010} we have described this theory in the framework of the so-called $k$-symplectic formalism \cite{aw1,gun,mt1,mt2}. In this section we consider the $k$-cosymplectic framework. Another attempts  in the framework of the multisymplectic formalism \cite{CIL1,{kt}} have been discussed in \cite{lmm2,{pau1},{pau2}}.

Along this section we only consider Hamiltonian systems defined on the phase-space $\R^k \times (T_k^1)^*Q$.

In Classical Field Theory the Hamilton-Jacobi equation is \cite{rund}
\begin{equation}\label{HJCFT}
    \derpar{W^\alpha}{x^\alpha} + H\Big(x^\beta, q^i, \derpar{W^\alpha}{q^i}\Big)=0
\end{equation}
where $W^1,\ldots, W^k\colon \rkq\to \r$.

The classical statement of time-dependent Hamilton-Jacobi equation for analytical mechanics is the following \cite{am}:
\begin{theorem}\label{t31}
    Let $H\colon \r\times T^*Q\to \r$ be a Hamiltonian and $T^*Q$ the symplectic manifold with the canonical symplectic structure $\omega=-d\theta$. Let $X_{H_t}$  be a Hamiltonian vector field on $T^*Q$ associated to the Hamiltonian  $H_t\colon T^*Q\to \r,\, H_t(\nu_q)=H(t,\nu_q)$, and $W\colon \r\times Q\to \r$ be a smooth function. The following two conditions are equivalent:
    \begin{enumerate}
        \item for every curve $c$ in $Q$ satisfying
            \[
                c'(t)=(\pi)_*\Big( X_{H_t}(dW_t(c(t)))\Big)
            \]
            the curve $t\mapsto dW_t(c(t))$ is an integral curve of $X_{H_t}$, where  $W_t\colon Q\to \r,\, W_t(q)=W(t,q)$.
        \item $W$ satisfies the Hamilton-Jacobi equation
            \[
                H(x,q^i,\derpar{W}{q^i}) + \derpar{W}{t}=\makebox{ constant on } T^*Q
            \]
            that is,
            \[
                H_t\circ dW_t+ \derpar{W}{t}= K(t)\,.
            \]
    \end{enumerate}
\end{theorem}

In this section we introduce a geometric version of the Hamilton-Jacobi theory based in the $k$-cosymplectic formalism. In the particular case $k=1$ we recover the above Theorem \ref{t31} for the time-dependent classical mechanics.

 For each $x=(x^1,\ldots, x^k)\in \rk$ we consider the following mappings
\[
    \begin{array}{lcl}
        \begin{array}{rccl}
         i_x\colon & Q & \to& \rk\times Q\\\noalign{\medskip}
          & q & \mapsto & (x,q)
        \end{array} & \quad and \quad & \begin{array}{rccl}
         j_x\colon & \tkqh & \to & \rktkqh\\\noalign{\medskip}
          & (\nu^1_q,\ldots, \nu^k_q) & \mapsto & (x,\nu^1_q,\ldots, \nu^k_q)
        \end{array}
    \end{array}
\]

Let $\gamma\colon \rkq\to \rktkqh$ be a section of $(\pi_Q)_{1,0}$. Let us observe that given a section $\gamma$ is equivalent to giving a mapping $\bar{\gamma}\colon \rkq\to \tkqh$.
If fact, given $\gamma$ we define $\bar{\gamma}=\bar{\pi}_2\circ \gamma$ where $\bar{\pi}_2$ is the canonical projection $\bar{\pi}_2\colon \rktkqh\to \tkqh$; conversely, given $\bar{\gamma}$ we define $\gamma$ as the composition $\gamma(x,q)=(j_x\circ \bar{\gamma})(x,q)$. Now, since $\tkqh$ is the Whitney sum of $k$ copies of the cotangent bundle, to give $\gamma$ is equivalent to give a family $(\bar{\gamma}^1,\ldots, \bar{\gamma}^k) $ of $1$-forms along the map $\pi_Q\colon \rkq\to Q$.

If we consider local coordinates $(x^\alpha, q^i, p^\alpha_i)$ we have the following local expressions:
\begin{equation}\label{localgamma}
    \begin{array}{l}
    \gamma(x^\alpha, q^i)= (x^\alpha, q^i, \gamma^\beta_j(x^\alpha, q^i))\,,\\\noalign{\medskip}
    \bar{\gamma}(x^\alpha, q^i) = (q^i, \gamma^\beta_j(x^\alpha, q^i))\,,\\\noalign{\medskip}
    \bar{\gamma}^\alpha(x,q)=\gamma^\alpha_j(x,q)dq^j(q)\,.
    \end{array}
\end{equation}


Moreover, along this section we suppose that each $\bar{\gamma}^\alpha$ satisfies that its exterior differential $d\bar{\gamma}^\alpha$ vanishes over two $\pi_{\rk}$-vertical vector fields. In local coordinates, using the local expressions (\ref{localgamma}), this condition implies that
\begin{equation}\label{gammacond}
    \derpar{\gamma^\alpha_i}{q^j}= \derpar{\gamma^\alpha_j}{q^i}\,.
\end{equation}

Now, let $Z=(Z_1,\ldots, Z_k)$ be a $k$-vector field on $\rktkqh$. Using $\gamma$ we can construct a $k$-vector field $Z^\gamma=(Z^\gamma_1,\ldots, Z^\gamma_k)$  on $\rkq$ such that the following diagram is commutative
\[
 \xymatrix{ \rktkqh
\ar@/^1pc/[dd]^{(\pi_Q)_{1,0}} \ar[rrr]^{Z}&   & &T_k^1(\rktkqh)\ar[dd]^{T^1_k(\pi_Q)_{1,0}}\\
  &  & &\\
 \rkq\ar@/^1pc/[uu]^{\gamma}\ar[rrr]^{Z^{\gamma}}&  & & T_k^1(\rkq) }
\]
that is,
$$Z^\gamma:= T^1_k(\pi_Q)_{1,0}\circ Z\circ \gamma\;.$$

Let us recall that for an arbitrary differentiable map $f:M\to N$, the induced map $T^1_kf:T^1_kM\to T^1_kN$ of $f$ is defined by
\begin{equation}\label{tkq: prolongation expr} T^1_kf({v_1}_x,\ldots, {v_k}_x)=(f_*(x)({v_1}_x),\ldots
,f_*(x)({v_k}_x)) \;,\end{equation}  where ${v_1}_x,\ldots,
{v_k}_x\in T_xM$, $x\in M$

Let us observe that if $Z$ is integrable then $Z^\gamma$ is also integrable.

In local coordinates, if each $Z_\alpha$ is locally given by
\[
    Z_\alpha= (Z_\alpha)_\beta\derpar{}{x^\beta} + Z^i_\alpha\derpar{}{q^i} + (Z_\alpha)^\beta_i\derpar{}{p^\beta_i}\,
\]
then $Z^\gamma_\alpha$ has the following local expression:
\begin{equation}\label{local z gamma kcos}
    Z^\gamma_\alpha= \big((Z_\alpha)_\beta\circ \gamma\big)\derpar{}{x^\beta} + (Z^i_\alpha\circ \gamma)\derpar{}{q^i}\,.
\end{equation}

In particular, if we consider the $k$-vector field $R=(R_1,\ldots, R_k)$ given by the Reeb vector fields, we obtain, by a similar procedure, a $k$-vector field $(R_1^\gamma,\ldots, R_k^\gamma)$ on $\rk\times Q$. In local coordinates,  since $R_\alpha=\nicefrac{\partial}{\partial x^\alpha}$ we have
\[
    R_\alpha^\gamma =\derpar{}{x^\alpha}\,.
\]

Next, we consider a Hamiltonian function $H\colon\rktkqh \longrightarrow \R$, and the corresponding Hamiltonian system on $\rktkqh$.
Notice that if $Z$ satisfies the Hamilton-De Donder-Weyl equations (\ref{geonah}), then we have
$$
(Z_\alpha)_\beta = \delta_{\alpha \beta}\,.
$$

\begin{theorem}[\bf Hamilton-Jacobi theorem]\label{HJthe}
 Let $Z\in \vf^k_H(\rktkqh)$ be a $k$-vector field solution to the $k$-cosymplectic Hamiltonian equation (\ref{geonah}) and $\gamma\colon \rkq\to \rktkqh$ be a  section of $(\pi_Q)_{1,0}$ with the property described above. If $Z$ is integrable then the following statements are equivalent:
    \begin{enumerate}
        \item If  a section $\psi\colon U\subset \rk\to \rkq$ of $\pi_{\rk}\colon \rkq\to \rk$ is an integral section of $Z^\gamma$ then $\gamma\circ \psi$ is a solution of the Hamilton-De Donder-Weyl equations (\ref{HEQ});
        \item $(\pi_Q)^*[d(H\circ \gamma\circ i_x)] + \sum_\alpha \, \imath_{R^\gamma_\alpha}d\bar{\gamma}^\alpha=0$ for any $x\in \rk$.
    \end{enumerate}
\end{theorem}
\proof

Let us suppose that a section $\psi\colon U\subset\rk\to \rk\times Q$ is an integral section of $Z^\gamma$. In local coordinates that means that if $\psi(x)=(x^\alpha,\psi^i(x))$, then
\[
    [(Z_\alpha)_\beta\circ \gamma](\psi(x))=\delta_{\alpha\beta},\quad (Z^i_\alpha\circ \gamma)(\psi(x))=\derpar{\psi^i}{x^\alpha}\,.
\]

Now by hypothesis, $\gamma\circ\psi\colon U\subset\rk\to \rktkqh$ is a solution of the Hamilton-De Donder-Weyl equation for $H$.  In local coordinates, if $\psi(x)=(x,\psi^i(x))$, then
 $\gamma\circ \psi(x)=(x,\psi^i(x),\gamma^\alpha_i(\psi(x)))$ and as it is a solution of the Hamilton-De Donder-Weyl equations for $H$, we have
\begin{equation}\label{HJkcos aux1}
\derpar{\psi^i}{x^\alpha}\Big\vert_{x} = \derpar{H}{p^\alpha_i}\Big\vert_{\gamma(\psi(x))} \makebox{ and } \ds\sum_{\alpha=1}^k\derpar{(\gamma^\alpha_i\circ \psi)}{x^\alpha}\Big\vert_{x} = -\derpar{H}{q^i}\Big\vert_{\gamma(\psi(x))}\,.
\end{equation}

Now, if we compute the differential of the function $H\circ \gamma\circ i_x\colon Q\to \r$, we obtain that:
\begin{equation}\label{HJ dif aux 2}
\begin{array}{ll}
    &(\pi_Q)^*[d(H\circ \gamma\circ i_x)] + \sum_\alpha \, \imath_{R^\gamma_\alpha}d\bar{\gamma}^\alpha\\\noalign{\medskip}
    =& \left(\derpar{H}{q^i}\circ \gamma\circ i_x + \left(\derpar{H}{p^\alpha_j}\circ \gamma\circ i_x\right)\left(\derpar{\gamma^\alpha_j}{q^i}\circ i_x\right) + \left(\derpar{\gamma^\alpha_i}{x^\alpha}\circ i_x\right)  \right)dq^i
    \end{array}\,.
\end{equation}

Therefore from (\ref{gammacond}), (\ref{HJkcos aux1}) and (\ref{HJ dif aux 2}) and taking into account that one can write $\psi(x)= (i_x\circ \pi_Q\circ \psi)(x)$, where $\pi_Q\colon \rkq \to Q$ is the canonical projection, we obtain
\[
    \begin{array}{rl}
       & ((\pi_Q)^*[d(H\circ \gamma\circ i_x)] + \sum_\alpha \, \imath_{R^\gamma_\alpha}d\bar{\gamma}^\alpha )(\pi_Q\circ\psi(x)) \\\noalign{\medskip}=& \left(\derpar{H}{q^i}\Big\vert_{\gamma(\psi(x))} + \derpar{H}{p^\alpha_j}\Big\vert_{\gamma(\psi(x))}\derpar{\gamma^\alpha_j}{q^i}\Big\vert_{\psi(x)} + \derpar{\gamma^\alpha_i}{x^\alpha}\Big\vert_{\psi(x)}
        \right)dq^i (\pi_Q\circ \psi(x))\\\noalign{\medskip}
        =& \left(- \ds\sum_{\alpha=1}^k\derpar{(\gamma^\alpha_i\circ \psi)}{x^\alpha}\Big\vert_{x} + \derpar{\psi^j}{x^\alpha}\Big\vert_{x} \derpar{\gamma^\alpha_j}{q^i}\Big\vert_{\psi(x)}  + \derpar{\gamma^\alpha_i}{x^\alpha}\Big\vert_{\psi(x)} \right) dq^i (\pi_Q\circ \psi(x))\\\noalign{\medskip}
         =& \left(- \ds\sum_{\alpha=1}^k\derpar{(\gamma^\alpha_i\circ \psi)}{x^\alpha}\Big\vert_{x} + \derpar{\psi^j}{x^\alpha}\Big\vert_{x} \derpar{\gamma^\alpha_i}{q^j}\Big\vert_{\psi(x)}    + \derpar{\gamma^\alpha_i}{x^\alpha}\Big\vert_{\psi(x)}\right) dq^i (\pi_Q\circ \psi(x))
         \\\noalign{\medskip}
         =&0\,.
    \end{array}
\]

As we have mentioned above, since $Z$ is integrable, the $k$-vector field $Z^\gamma$ is also integrable, and then for each point $(x,q)\in \rkq$ we have an integral section $\psi\colon U\subset \rk\to \rk\times Q$ of $Z^\gamma$ passing trough this point. Therefore, for any $x\in \rk$, we get
\[
    (\pi_Q)^*[d(H\circ \gamma\circ i_x)]+ \sum_\alpha \, \imath_{R^\gamma_\alpha}d\bar{\gamma}^\alpha=0\,.
\]

Conversely, let us suppose that $(\pi_Q)^*[d(H\circ \gamma\circ i_x)]+ \sum_\alpha \, \imath_{R^\gamma_\alpha}d\bar{\gamma}^\alpha=0$ and take $\psi$ an integral section of $Z^\gamma$. We now will prove that $\gamma\circ \psi$ is a solution to the Hamilton-De Donder-Weyl field equations for $H$.

Since $(\pi_Q)^*[d(H\circ \gamma\circ i_x)]+ \sum_\alpha \,  \imath_{R^\gamma_\alpha}d\bar{\gamma}^\alpha=0$ from (\ref{HJ dif aux 2}) we obtain
\begin{equation}\label{HJ dif aux 3}
    \derpar{H}{q^i}\circ \gamma\circ i_x + \left(\derpar{H}{p^\alpha_j}\circ \gamma\circ i_x\right)\left(\derpar{\gamma^\alpha_j}{q^i}\circ i_x\right) + \left(\derpar{\gamma^\alpha_i}{x^\alpha}\circ i_x\right)  =0\,.
\end{equation}

From (\ref{partials h X}) and (\ref{local z gamma kcos}) we know that
\begin{equation}\label{zgammaH}
    Z^\gamma_\alpha= \derpar{}{x^\alpha} + \Big(\derpar{H}{p^\alpha_i}\circ \gamma\Big)\derpar{}{q^i}\,,
\end{equation}
and then since $\psi(x,q)=(x,\psi^i(x,q))$ is an integral section of $Z^\gamma$ we deduce that
\begin{equation}\label{HJ kcos aux 4}
    \derpar{\psi^i}{x^\alpha}= \derpar{H}{p^\alpha_i}\circ \gamma\circ \psi\,.
\end{equation}

On the other hand, from (\ref{gammacond}), (\ref{HJ dif aux 3}) and (\ref{HJ kcos aux 4}) we get
\[
    \begin{array}{ll}
       & \ds\sum_{\alpha=1}^k\derpar{(\gamma^\alpha_i\circ \psi)}{x^\alpha}\Big\vert_{x}= \ds\sum_{\alpha=1}^k\left(\derpar{\gamma^\alpha_i}{x^\alpha}\Big\vert_{\psi(x)} +\derpar{\gamma^\alpha_i}{ q^j}\Big\vert_{\psi(x)}\derpar{\psi^j}{x^\alpha}\Big\vert_{x}\right) = \\\noalign{\medskip} =& \ds\sum_{\alpha=1}^k\left(\derpar{\gamma^\alpha_i}{x^\alpha}\Big\vert_{\psi(x)} +\derpar{\gamma^\alpha_i}{ q^j}\Big\vert_{\psi(x)}\derpar{H}{p^\alpha_j}\Big\vert_{\gamma(\psi(x))}\right)
       \\\noalign{\medskip} =&  \ds\sum_{\alpha=1}^k\left(\derpar{\gamma^\alpha_i}{x^\alpha}\Big\vert_{\psi(x)} +\derpar{\gamma^\alpha_j}{ q^i}\Big\vert_{\psi(x)}\derpar{H}{p^\alpha_j}\Big\vert_{\gamma(\psi(x))}\right) = -\derpar{H}{q^i}\Big\vert_{\gamma(\psi(x))}
    \end{array}
\]
and thus we have proved that $\gamma\circ \psi$ is a solution to the Hamilton-de Donder-Weyl equations.
\qed

\begin{theorem}
    Let $Z\in \vf^k_H(\rktkqh)$ be a $k$-vector field solution to the $k$-cosymplectic Hamiltonian equation (\ref{geonah}) and $\gamma\colon \rkq\to \rktkqh$ be a  section of $(\pi_Q)_{1,0}$ satisfying the same conditions of the above theorem. Then, the following statements are equivalent:
    \begin{enumerate}
        \item $Z\vert_{Im\,\gamma} - T^1_k\gamma(Z^\gamma)\in \ker \omega^\sharp\cap \ker \eta^\sharp$, being $\omega^\sharp$ and $\eta^\sharp$ the vector bundle morphism defined in Remark \ref{existence kcos sol};
        \item $(\pi_Q)^*[d(H\circ \gamma\circ i_x)] + \sum_\alpha \, \imath_{R^\gamma_\alpha}d\bar{\gamma}^\alpha=0$.
    \end{enumerate}
\end{theorem}

\proof A direct computation shows that $Z_\alpha\vert_{Im\,\gamma} - T\gamma(Z_\alpha^\gamma)$ has the following local expression
\[
    \left((Z_\alpha)^\beta_j\circ \gamma - \derpar{\gamma^\beta_j}{x^\alpha} - (Z^i_\alpha\circ \gamma)\derpar{\gamma^\beta_j}{q^i} \right) \derpar{}{p^\beta_j}\circ \gamma\,.
\]

Thus from (\ref{kerkco}) we know that $Z\vert_{Im\,\gamma} - T^1_k\gamma(Z^\gamma)\in \ker \omega^\sharp\cap \ker \eta^\sharp$ if and only if
\begin{equation}\label{condker}
 \ds\sum_{\alpha=1}^k\left((Z_\alpha)^\alpha_j\circ \gamma - \derpar{\gamma^\alpha_j}{x^\alpha} - (Z^i_\alpha\circ \gamma)\derpar{\gamma^\alpha_j}{q^i} \right)=0\,.
\end{equation}

Now we are ready to prove the result.

Assume that $(i)$ holds, then from (\ref{partials h X}), (\ref{gammacond}) and (\ref{condker}) we obtain
\[
\begin{array}{ll}
    0=& \ds\sum_{\alpha=1}^k\left((Z_\alpha)^\alpha_j\circ \gamma - \derpar{\gamma^\alpha_j}{x^\alpha} - (Z^i_\alpha\circ \gamma)\derpar{\gamma^\alpha_j}{q^i} \right)\\\noalign{\medskip}
    =& -\left(\left(\derpar{H}{q^j}\circ \gamma\right) + \sum_\alpha \, \derpar{\gamma^\alpha_j}{x^\alpha} + \left(\derpar{H}{p^\alpha_i}\circ \gamma\right)\derpar{\gamma^\alpha_j}{q^i} \right)\\\noalign{\medskip}
    =& -\left(\left(\derpar{H}{q^j}\circ \gamma\right) + \sum_\alpha \, \derpar{\gamma^\alpha_j}{x^\alpha} + \left(\derpar{H}{p^\alpha_i}\circ \gamma\right)\derpar{\gamma^\alpha_i}{q^j} \right)\,.
\end{array}
\]
Therefore $(\pi_Q)^*[d(H\circ \gamma\circ i_x)] + \imath_{R^\gamma_\alpha}d\bar{\gamma}^\alpha=0$ (see (\ref{HJ dif aux 2})).

The converse is proved in a similar way by reversing the arguments.
\qed

\begin{corollary}
    Let $Z\in \vf^k_H(\rktkqh)$ be a  solution of (\ref{geonah}) and $\gamma\colon \rkq\to \rktkqh$ be a  section of $(\pi_Q)_{1,0}$ as in the above theorem. If $Z$ is integrable then the following statements are equivalent:
    \begin{enumerate}
        \item $Z\vert_{Im\gamma} - T^1_k\gamma(Z^\gamma)\in \ker \omega^\sharp\cap \ker \eta^\sharp$;
        \item $(\pi_Q)^*[d(H\circ \gamma\circ i_x)] + \sum_\alpha \, \imath_{R^\gamma_\alpha}d\bar{\gamma}^\alpha=0$;
        \item If  a section $\psi\colon U\subset \rk\to \rkq$ of $\pi_{\rk}\colon \rkq\to \rk$ is an integral section of $Z^\gamma$ then $\gamma\circ \psi$ is a solution of the Hamilton-De Donder-Weyl equations (\ref{HEQ}).
    \end{enumerate}
\end{corollary}

Let us observe that there exist $k$ local functions $W^\alpha$ such that $\bar{\gamma}^\alpha=dW^\alpha_x$ being $W_x$ the function defined by $ W^\alpha_x(q)=W^\alpha(x,q)$. Thus $\gamma^\alpha_i=\nicefrac{\partial W^\alpha}{\partial q^i}$ (see \cite{KN}). Therefore, the condition
\[
(\pi_Q)^*[d(H\circ \gamma\circ i_x)]+ \sum_{\alpha}\imath_{R^\gamma_\alpha}d\bar{\gamma}^\alpha=0
\]
can be equivalently written as
\[
\derpar{}{q^i}\left(\derpar{W^\alpha}{x^\alpha} + H(x^\beta,q^i,\derpar{W^\alpha}{q^i})\right)=0.
\]

The above expressions mean that
\[
    \derpar{W^\alpha}{x^\alpha} + H(x^\beta,q^i,\derpar{W^\alpha}{q^i})=K(x^\beta)
\]
so that if we put $\tilde{H}=H-K$ we deduce the standard form of the Hamilton-Jacobi equation (since $H$ and $\tilde{H}$ give the same Hamilton-De Donder-Weyl equations).
\begin{equation}\label{HJeq W}
    \derpar{W^\alpha}{x^\alpha} + \tilde{H}(x^\beta,q^i,\derpar{W^\alpha}{q^i})=0\,.
\end{equation}

Therefore the equation
\begin{equation}\label{HJkcoseq}
(\pi_Q)^*[d(H\circ \gamma\circ i_x)]+ \sum_{\alpha} \imath_{R^\gamma_\alpha}d\bar{\gamma}^\alpha=0
\end{equation}
can be considered as a geometric version of the Hamilton-Jacobi equation for $k$-cosymplectic field theories.

\section{An example}\label{example}
In this section we will apply our method to a particular example in classical field theories.


The equation of a scalar field $\phi$ (for instance the gravitational field) which acts on the $4$-dimensional space-time is (see \cite{kt}):
                    \begin{equation}\label{scalar}
                        (\square + m^2)\phi =F'(\phi)\,,
                    \end{equation}
               where $m$ is the mass of the particle over which the fields acts, $F$ is a scalar function such that $F(\phi)-\ds\frac{1}{2}m^2\phi^2$ is the potential energy of the particle of mass $m$, and $\square$ is the Laplace-Beltrami operator given by
                    \[
                        \square\phi\colon = div\, grad \phi = \ds\frac{1}{\sqrt{-g}}\derpar{}{x^\alpha}\left(\sqrt{-g}g^{\alpha\beta}\derpar{\phi}{x^\beta}\right)\,,
                    \]
               $(g_{\alpha\beta})$ being a pseudo-Riemannian metric tensor in the $4$-dimensional space-time of signature $(-+++)$, and $\sqrt{-g}=\sqrt{-\det{g_{\alpha\beta}}}$.

We consider the Lagrangian
\[
L(x^1,x^2,x^3,x^4, q, v_1,v_2,v_3, v_4)= \sqrt{-g}\Big(F(q)- \ds\frac{1}{2}m^2q^2\Big)+\ds\frac{1}{2}g^{\alpha\beta}v_\alpha v_\beta\,,
\]
where $q$ denotes the scalar field $\phi$ and $v_\alpha$ the partial derivative $\nicefrac{\partial \phi}{\partial x^\alpha}$. Then the equation (\ref{scalar}) is just the Euler-Lagrange equation associated to $L$.

               Consider the Hamiltonian function $H\in\mathcal{C}^\infty(\r^4\times (T^1_4)^*\r)$ given by
                    \[
                        H(x^1,x^2,x^3,x^4,q, p^1, p^2, p^3, p^4)=  \ds\frac{1}{2\sqrt{-g}}g_{\alpha\beta}p^\alpha p^\beta-\sqrt{-g}\left( F(q)-\ds\frac{1}{2}m^2q^2\right)\,,
                    \]
               where $(x^1,x^2,x^3,x^4)$ are the coordinates on $\r^4$, $q$ denotes the scalar field $\phi$ and $(x^1,x^2,x^3,x^4,q,p^1,p^2,p^3,p^4)$ the canonical coordinates on $\r^4\times (T^1_4)^*\r$. Let us recall that this Hamiltonian function can be obtained from the Lagrangian $L$ just using the Legendre transformation defined in \cite{merino3,arxive}.

Then
                \begin{equation}\label{partial Ham scalar k-co}
                    \derpar{H}{q}=-\sqrt{-g}\Big(F'(q)-m^2q\Big),\quad \derpar{H}{p^\alpha}=\frac{1}{\sqrt{-g}}g_{\alpha\beta}p^\beta\,.
                \end{equation}

The Hamilton-Jacobi equation becomes
\begin{equation}\label{campo escalar HJ}
    -\sqrt{-g}\Big(F'(q)-m^2q\Big)+ \ds\frac{1}{\sqrt{-g}}g_{\alpha\beta}\gamma^\beta\derpar{\gamma^\alpha}{q} + \ds\derpar{\gamma^\alpha}{x^\alpha}=0\,.
\end{equation}

Since our main goal is to show how the method developed in Section 3 works, we will consider, for simplicity, the following particular case:
$$
F(q)=\frac{1}{2}m^2q^2,
$$
being $(g_{\alpha\beta})$ the Minkowski metric on $\r^4$, i.e. $(g_{\alpha\beta})=diag (-1, 1,1,1)$.

Let $\gamma\colon \r^4\to \r^4\times (T^1_k)^*\r$ be the section of $(\pi_\r)_{1,0}$ defined by the family of $4$ $1$-forms along of $\pi_\r\colon\r^4\times \r\to \r$
\[
    \bar{\gamma}^\alpha= \frac{1}{2}C_\alpha q^2dq\,
\]
with $1\leq \alpha\leq 4$ and where $C_\alpha$ are four constants such that $C_1^2=C_2^2+C_3^3+C_4^2$.  This section $\gamma$ satisfies the Hamilton-Jacobi equation (\ref{campo escalar HJ}) that in this particular case is given by
\[
 -\frac{1}{2} C_1^2q^3 + \ds\frac{1}{2}\sum_{a=2}^4C_a^2q^3=0\,,
\]
therefore, the condition $(ii)$ of the Theorem \ref{HJthe} holds.

The $4$-vector field $Z^\gamma=(Z^\gamma_1,Z^\gamma_2,Z^\gamma_3,Z^\gamma_4)$ is locally given by
\[
    Z^\gamma_1=\derpar{}{x^1}-\frac{1}{2}C_1 q^2\derpar{}{q}\,,\quad Z^\gamma_a=\derpar{}{x^a}+\frac{1}{2}C_a q^2\derpar{}{q}\,,
\]
with $a=2,3,4$. The map $\psi\colon\r^4\to\r^4\times \r$ defined by
\[
    \psi(x^1,x^2,x^3,x^4)=\ds\frac{2}{C_1x^1-C_2x^2-C_3x^3-C_4x^4+C}\,,\quad C\in \r\,,
\]
 is an integral section of the $4$-vector field $Z^\gamma$.

 By Theorem \ref{HJthe} one obtains that the map $\varphi=\gamma\circ \psi$, locally given by
 \[
    (x^\alpha)\to (x^\alpha,\psi(x^\alpha), \frac{1}{2}C_\beta(\psi(x^\alpha))^2)\,,
 \]
 is a solution of the Hamilton-De Donder-Weyl equations associated to $H$, that is,
 \[
 \begin{array}{rcl}
    0&=&\ds \sum_{\alpha=1}^4\derpar{}{x^\alpha}\Big(\frac{1}{2} C_\alpha\psi^2\Big)\,,\\\noalign{\medskip}
    -\frac{1}{2}C_1\psi^2 &=&\derpar{\psi}{x^1}\,,\\\noalign{\medskip}
    \frac{1}{2}C_a\psi^2&=&\derpar{\psi}{x^a}\,,\quad a=2,3,4\,.
     \end{array}
 \]
Let us observe that  these equations imply that the scalar field $\psi$ is a solution to the $3$-dimensional wave equation.

In this particular example the functions $W^\alpha$ are given by
\[
W^\alpha(x,q)=\ds\frac{1}{6}C_\alpha q^3 + h(x)\,,
\]
where $h\in \mathcal{C}^\infty (\r^4)$.

\bigskip

In \cite{pau2, ZS}, the authors describe an alternative method that can be compared with the above one.

First, we consider the set of functions $W^\alpha\colon \r^4\times \r\to \r, \, 1\leq \alpha\leq 4$ defined by
\[
    W^\alpha(x,q)= (q-\ds\frac{1}{2}\phi(x))\sqrt{-g}g^{\alpha\beta}\derpar{\phi}{x^\beta}\,,
\]
where $\phi$ is a solution to the Euler-Lagrange equation (\ref{scalar}).
Using these functions we can consider a section $\gamma$ of $(\pi_\r)_{1,0}\colon \r^4\times (T^1_4)^*\r\to \r^4\times\r$  with components
\[
    \gamma^\alpha= \derpar{W^\alpha}{q} = \sqrt{-g}g^{\alpha\beta}\derpar{\phi}{x^\beta}\,.
\]

By a direct computation we obtain that this section $\gamma$ is a solution to the Hamilton-Jacobi equation (\ref{HJkcoseq}).

Now from (\ref{zgammaH}) and (\ref{partial Ham scalar k-co})  we obtain the $4$-vector field $Z^\gamma$ is given by
\begin{equation}\label{examplezgamma}
           Z^\gamma_\alpha = \derpar{}{x^\alpha} +\derpar{\phi}{x^\alpha}\derpar{}{q}\,.
\end{equation}

Let us observe that $Z^\gamma$ is an integrable $4$-vector field on $\r^4\times \r$. Using the Hamilton-Jacobi theorem we obtain that  $\sigma=(id_{\r^4},\phi)\colon \r^4\to \r^4\times \r$ is an integral section of the $4$-vector field $Z^\gamma$ defined by (\ref{examplezgamma}), then $\gamma\circ \sigma$ is  a solution of the Hamilton-De Donder Weyl equation associated with the Hamiltonian of the massive scalar field.

If we now consider the particular case $F(q)=m^2q^2$, we obtain the Klein-Gordon equation; this is just the case discussed in \cite{pau2}.

\section*{Acknowledgments}
This work has been partially supported by {\sl Ministerio de
Ciencia e Innovaci\'{o}n}, Projects MTM2010-21186-C02-01, MTM2011-22585, MTM2011-15725-E; the European project IRSES-project GeoMech-246981; the ICMAT Severo Ochoa project SEV-2011-0087 and {\sl Gobierno de Arag\'{o}n} E24/1.


\begin{thebibliography}{99}

\bibitem{am} R.A. Abraham, J.E. Marsden: {Foundations of Mechanics}
(Second Edition), Benjamin-Cummings Publishing Company, New York,
1978.

\bibitem{aw1} A. Awane: $k$-symplectic structures. {\it J. Math. Phys.} {\bf 33} (1992), 4046--4052.
%

\bibitem{BLM-2012} M. Barbero-Li\~{n}an, M. de Le\'{o}n, D. Mart\'{\i}n de Diego: Lagrangian submanifolds and Hamilton-Jacobi equation. 	 arXiv:1209.0807 [math-ph]

\bibitem{bertin} M.C. Bertin, B.M. Pimentel, P.J. Pompeia:
Hamilton-Jacobi approach for first order actions and theories with higher order derivatives.
{\it Ann. Physics} {\bf 323} (2008), no. 3, 527--547.

\bibitem{bruno} D. Bruno: Constructing a class of solutions for the Hamilton-Jacobi equations in field theory.
{\it J. Math. Phys.} {\bf 48}, 112902 (2007), no. 11.

\bibitem{CIL1} F. Cantrijn, A. Ibort, M. de Le\'on: On the geometry of multisymplectic manifolds.
{\it J. Austral. Math. Soc. (Series A)\/} {\bf 66} (1999), 303--330.

\bibitem{CGMMR} J. F. Cari\~{n}ena, X. Gr\`{a}cia, G. Marmo, E. Mart\'{\i}nez,
M.C. Mu\~{n}oz-Lecanda, N. Rom\'{a}n-Roy: Geometric Hamilton-Jacobi
theory. {\it Int. J. Geom. Methods Mod. Phys.} \textbf{3} (2006),
no. 7, 1417--1458.

\bibitem{pepin2} J. F. Cari\~{n}ena, X. Gr\`{a}cia, G. Marmo, E. Mart\'{\i}nez,
M.C. Mu\~{n}oz-Lecanda, N. Rom\'{a}n-Roy: Geometric Hamilton-Jacobi Theory for Nonholonomic Dynamical Systems.
{\it Int. J. Geom. Methods Mod. Phys.} \textbf{7} (2010),
no. 3, 431--454.

\bibitem{gun} C. G\"{u}nther: The polysymplectic Hamiltonian
formalism in field theory and calculus of variations I: The local
case. {\it J. Differential Geom.} {\bf 25} (1987), no. 1, 23-53.

\bibitem{ultimo}
{H. Loumu-Fergane, A. Belaidi}: Multisymplectic geometry and $k$-cosymplectic structure for the field theories and the relativistic mechanics.
{\it  Int. J. Geom. Methods Mod. Phys}, {\bf 10} (2013), no. 4,
(doi: 10.1142/S0219887813500011)

\bibitem{kt}
{J. Kijowski, W. M. Tulczyjew}:  A symplectic framework for
field theories. {\it Lecture Notes in Physics},  107. Springer-Verlag,
Berlin-New York, 1979.

\bibitem{KN}
{S. Kobayashi, K. Nomizu}: Foundations of differential geometry. {\it Interscience Publishers} vol I (1963).

\bibitem{lmm1} M. de Le\'on, D. Iglesias-Ponte, D. Mart\'{\i}n de Diego:
Towards a Hamilton-Jacobi theory for nonholonomic mechanical
systems. {\it  J. Phys. A}  {\bf 41}  (2008),  no. 1, 015205, 14 pp

\bibitem{lmm2} M. de Le\'on, J.C. Marrero, D. Mart\'{\i}n de Diego: A geometric Hamilton-Jacobi theory for classical field theories. In {\it Variations, Geometry and Physics, in honour of Demeter Krupka's sixty-fifth birthday}, O. Krupkova and D. J. Saunders (Editors), Nova Science Publishers Inc., New York 2009, pp. 129--140.


\bibitem{lmm3} M. de Le\'on, J.C. Marrero, D. Mart\'{\i}n de Diego:
Linear almost Poisson structures and Hamilton-Jacobi equation. Applications to nonholonomic Mechanics. {\it J. Geom. Mech.} \textbf{2} (2010), no. 2, 159--198

\bibitem{LMMSV-2010}
M. de Le\'on, J.C. Marrero, D. Mart\'{\i}n de Diego, M. Salgado, S. Vilari\~{n}o: Hamilton-Jacobi theory in $k$-symplectic field theories. {\it Int. J. Geom. Methods Mod. Phys.} 7 (2010), no. 8, 1491--1507.

\bibitem{mt1} M. de Le\'{o}n, I. M\'endez, M. Salgado: $p$-almost tangent
structures. {\it Rend.  Circ. Mat. Palermo} Serie II {\bf XXXVII}
(1988), no. 2, 282-294.

\bibitem{mt2} M. de Le\'{o}n, I. M\'{e}ndez, M. Salgado:
Integrable $p$-almost tangent structures and tangent bundles of
$p^1$-ve\-lo\-ci\-ties. {\it Acta Math. Hungar.} {\bf 58}(1-2)
(1991), 45-54.

\bibitem{merino1} M. de Le\'on, E. Merino, M. Salgado:
$k$-cosymplectic manifolds and Lagrangian field theories.
{\it J. Math. Phys.} 42 (2001), no. 5, 2092--2104.

\bibitem{merino2} M. de Le\'on, E. Merino, M. Salgado:
Stable almost cotangent structures. {\it Boll. Un. Mat. Ital. B} (7) 11 (1997), no. 3, 509--529.

\bibitem{merino3} M. de Le\'on, E. Merino, J.A. Oubi\~{n}a, P.R. Rodrigues, M. Salgado:
Hamiltonian systems on $k$-cosymplectic manifolds.
{\it J. Math. Phys.} 39 (1998), no. 2, 876--893.

\bibitem{arxive} M. de Le\'on, M. McLean, L.K. Norris, A.M. Rey, M. Salgado:
Geometric Structures in Field Theory. ArXiv:math-ph/0208036v1
(2002).

\bibitem{mor} A. Morimoto: Liftings of some types of tensor fields and
connections to tangent $p^r$-velocities. {\it Nagoya Math. J.} {\bf
40} (1970), 13-31.

\bibitem{MRS-2004}
F. Munteanu, A. M. Rey, M. Salgado: The G\"{u}nther's
formalism in classical field theory: momentum map and reduction. {\it
J. Math. Phys.} {\bf 45} (2004), no. 5, 1730--1751.

\bibitem{pau1} C. Paufler, H. Romer: De Donder-Weyl equations and
multisymplectic geometry. In: {\it XXXIII Symposium on Mathematical Physics (Tor\'un, 2001).
Rep. Math. Phys.}  {\bf 49}  (2002), no. 2-3, 325--334.

\bibitem{pau2} C. Paufler, H. Romer: Geometry of Hamiltonian $n$-vector fields
in multisymplectic field theory. {\it J. Geom. Phys.} {\bf 44}  (2002), no. 1, 52--69.

\bibitem{RRSV-2011} N. Rom\'{a}n-Roy, M.A. Rey, M. Salgado, S. Vilari\~{n}o: On the k-symplectic, k-cosymplectic and multisymplectic formalisms of classical field theories. {\it J. Geom. Mech.} {\bf 3} (2011), no. 1, 113--137.

\bibitem{rosen} G. Rosen: Hamilton-Jacobi functional theory for the integration of classical field equations.
{\it Internat. J Theoret. Phys.} {\bf 4} (1971), 281--285.

\bibitem{rund} H. Rund: The Hamilton-Jacobi Theory in the Calculus of
Variations. Hazell, Watson and Viney Ltd., Aylesbury,
Buckinghamshire, U.K. 1966.


\bibitem{vita} L. Vitagliano: The Hamilton-Jacobi formalism for higher order field theories. {\it Int. J. Geom. Methods Mod. Phys.} 7 (2010), no. 8, 1413--1436.

\bibitem{ZS}
Zhi-Qiang Guo, I. Schmidt: Converting Classical Theories to Quantum Theories by Solutions of the Hamilton-Jacobi Equation. 2012. Preprint. arXiv:1204.1361v2 [hep-th].
\end{thebibliography}
\end{document}